\documentclass[twocolumn]{jpsj3}
\usepackage{txfonts}
\usepackage{cite}
\usepackage{epsf}
\usepackage{epsfig}
\usepackage{epstopdf}
\usepackage{float}
\usepackage{amssymb}
\usepackage{amsfonts}
\usepackage{color}
\usepackage{multicol}

\def\om{{\omega}}

\def\g{{\bf{g}}}

\def\beq{\begin{equation}}
\def\eeq{\end{equation}}
\def\beqa{\begin{eqnarray}}
\def\eeqa{\end{eqnarray}}

\def\g0{{\gamma_0}}

\def\prb{Phys.\ Rev.\ B }
\def\prl{Phys.\ Rev.\ Lett.\ }
\def\jpcm{J.\ Phys.\ Condens.\ Matter }

\def\etal{{\it et.\ al} }

\def\RMP{Rev.\ Mod.\ Phys.\ }

\newcommand\bear{\begin{eqnarray}}
\newcommand\eear{\end{eqnarray}}
\newcommand\bea{\begin{align}}
\newcommand\ena{\end{align}}
\usepackage{comment}
\usepackage{setspace}
\usepackage{amsmath,amssymb,bm}
\usepackage{color}

\def\CPA{\scriptscriptstyle CPA}
\title{Random local attraction driven metal-superconductor transitions}
\author{Naushad Ahmad Kamar  and N. S. Vidhyadhiraja \thanks{E-mail- raja@jncasr.ac.in}}
\inst{Theoretical Sciences Unit\\Jawaharlal Nehru Centre for Advanced Scientific Research,\\
 Jakkur, Bangalore 560 064, India.}
\abst{
In this paper, we  investigate the disordered attractive Hubbard model
by combining dynamical mean field theory, coherent potential approximation and iterated perturbation theory for superconductivity as an impurity solver.
Disorder is introduced in the local attraction $U$. We assume that  $U$ is distributed according to a bimodal probability distribution, wherein an $x$ fraction of sites are pairing centers ($U\neq0$) and $(1-x)$  fraction of sites  are non-interacting ($U=0$). It is found that beyond a critical $x=x_c$,
a first order metal-superconductor phase transition leads to 
superconductivity being induced in the interacting as well as non-interacting sites.}

\kword{Disordered s-wave superconductors, Metal- superconductor transitions, dynamical mean field theory }
\begin{document}
\maketitle
\section{Introduction}
The combined effect of disorder and correlations on the superconducting state has been
extensively studied since many decades but a complete picture has not emerged yet~\cite{ 3PA, 3Belitz }. A scenario of randomly located pairing centers leading
to $s$-wave superconductivity may not be far-fetched and might be appropriate
 for real systems such as Tl-doped PbTe.
The semiconducting PbTe, when doped with Tl (Pb$_{1-x}$Tl$_x$Te), exhibits a superconducting
ground state~\cite{ 3Matsushita1, 3Matsushita2} beyond a critical concentration ($x = x_c =0.3\%$) of Tl.  The consensus regarding the origin of such doping induced 
superconductivity is that the Tl dopants represent spatially inhomogeneous centers of negative attractive interactions ($-U$), which nucleate Cooper pairing, and hence
lead to superconductivity~\cite{3Micnas}. 
It was found~~\cite{3Matsushita1, 3Matsushita2} that the transition temperature, $T_c$, increases with increasing $x$ thus supporting the idea that Tl-dopants nucleate superconductivity.


A spatial distribution of attractive interaction centers may be mathematically represented
by the attractive Hubbard model (AHM), with random attractive Hubbard interactions.  We use the framework of dynamical mean field
 theory (DMFT)~\cite{3Antoine, 3Kotliar, 3Dieter} combined with coherent potential approximation (CPA)~\cite{3Elliott, 3Jannis} to investigate this model. We employ two impurity solvers within DMFT:
 iterated perturbation theory for superconducting case (IPTSC)~\cite{3Arti} and static mean-field theory~\cite{3Shenoy, 3Litak}. The random interaction is taken to be distributed according to a bimodal probability distribution. While $x$ fraction of sites have
an attractive interaction, $-U$, $1-x$ fraction of sites are non-interacting. It is observed that beyond a critical $x=x_c$ the system is superconducting, but for large $U$  values, a small value of $x$ is sufficient to make whole system superconducting. 
The clean limit of the AHM has been
extensively studied using Bogoliubov-de Gennes type mean field (BdGMF) theories and more recently using iterated perturbation theory with superconducting bath (IPTSC)~\cite{3Arti, 3Naushad1, 3Naushad2}, numerical renormalization group (NRG)~\cite{3Bauer} and continuous time
quantum Monte Carlo (CTQMC)~\cite{3Werner} within DMFT. The main issue that has been focused upon is the BCS-BEC crossover for different fillings and interaction strengths. The IPTSC method 
is known to benchmark excellently when compared to results
from numerical renormalization group for the bulk
AHM~\cite{3Bauer}.
In this paper, we carry out
a detailed study of the  AHM with inhomogeneous interaction
 by combining CPA with DMFT and iterated perturbation theory for superconductivity(IPTSC). To distinguish between
dynamical and static effects, we have also carried out BdGMF studies within CPA+DMFT.
This paper is structured as
follows: In the following section, we outline the model and the formalism used. Next, we 
present our results for the local order parameter, spectra and disorder induced phase transitions. We conclude in the final section.
\section{Model and Method}
 We consider the single band attractive Hubbard model(AHM), which is given by the following Hamiltonian,
\begin{eqnarray}
{\cal H}=
\sum_{i\sigma}\epsilon c^{\dagger}_{i\sigma}c^{\phantom{\dagger}}_{i\sigma}-t\sum_{\langle ij\sigma\rangle} \left(c^{\dagger}_{i\sigma}c^{\phantom{\dagger}}_{j\sigma}+h.c\right)- \nonumber  \\ \sum_{i} |U_i|\left(n_{i\uparrow}-\frac{1}{2}\right)  \left(n_{i\downarrow }-\frac{1}{2}\right)- 
\mu\sum_{i\sigma}c^{\dagger}_{i\sigma}c^{\phantom{\dagger}}_{i\sigma}
\label{eq:3eq1}
\end{eqnarray}
Where ${c}_{i\sigma}$ annihilates an electron on $i^{\rm th}$ lattice site with spin $\sigma$, and ${n}_{i\sigma}=c^{\dagger}_{i\sigma}c^{\phantom{\dag}}_{i\sigma}$, t is nearest neighbour hopping matrix, ${\epsilon}$ is onsite energy, and $\mu$ is chemical potential. The local disorder is given by random attractive Hubbard interaction, which is distributed according to the bimodal probability distribution function $P_U(U_{i})$ 
\begin{equation}
{P_U(U_{i})}=(1-x)\delta(U_{i})+x\delta(U_{i}+U)\,,
\label{eq:3eq2}
\end{equation}
where $1-x$ and $x$ are fractions of lattice sites with interaction $U_i=0$ and $U_i=-U$, respectively.
Within DMFT, the lattice model is mapped onto a single-impurity model
embedded in a self-consistently determined bath. For the present problem,
the effective medium is in a superconducting state, hence the Nambu
formalism must be used.
The effective action~\cite{3Antoine, 3Naushad} for a given site $i$ within DMFT in Nambu formalism is given by
\begin{eqnarray*}
\hspace*{0cm} \hat{S}_{eff}(i)=
-\int_{0}^{\beta}\ d\tau_1 \int_{0}^{\beta}\ d\tau_2 \Psi^{\dagger}_{i}(\tau_1)\hat {\cal G}^{-1}(\tau_1-\tau_2)\Psi_{i}(\tau_2)-\\|U_i|\int_{0}^{\beta}d\tau n_{i\uparrow}(\tau)n_{i\downarrow}(\tau)
\end{eqnarray*}
where $\Psi_i(\tau)$, the two component Nambu's spinor and $\hat {\cal G}(\tau)$, the host Green's function in Nambu formalism are given by
 \begin{eqnarray*}
\hspace*{0cm}  \Psi_i(\tau)  =
 \left[ \begin{array}{ll}
{ c_{i\uparrow}(\tau)} \\
 { c^{\dagger}_{i\downarrow}}(\tau) 
\end{array} \right]
\end{eqnarray*}
and
 \begin{equation*}
\hspace*{0cm} {\hat {\cal G}}(\omega) =
 \left[ \begin{array}{ll}
{ \omega^{+}-\epsilon+\mu -\Delta_{11}(\omega)} & { -\Delta_{12}(\omega)} \\
 {-\Delta_{21}(\omega) }
 & {\omega^{+}+\epsilon-\mu-\Delta_{22}(\omega)}
\end{array} \right]^{-1}
\end{equation*}
Where $\om^{+}=\om+i\eta$, and $\eta\rightarrow 0^+$. 
The impurity Green's function in Nambu formalism is given as
\begin{equation*}
\hat{G}^{i}(\tau)=-\langle T_{\tau}\Psi_i(\tau)\Psi^{\dagger}_i(0)\rangle
\end{equation*}
 To discuss how disorder affects superconductivity, we  use  CPA with DMFT. The impurity Green's function in Nambu formalism is given as 
 \begin{eqnarray}
\hspace*{0cm} {\hat{G}^{i}}(\omega) =
 \left[ \begin{array}{ll}
{ \xi_{11}(\om)-\Sigma_{i}(\omega) } &  { -\Delta_{12}(\omega)-S_{i}(\omega)} \\
 { -\Delta_{21}(\omega)-S_{i}(\omega) }
 &  {\xi_{22}(\om)+\Sigma^{*}_{i}(-\omega)}
\end{array} \right]^{-1} 
 \label{eq:3eq3}
\end{eqnarray}
where, $\xi_{11}(\om)=\omega^{+}+\mu-\epsilon-\Delta_{11}(\omega)$, \ $\xi_{22}(\om)=\omega^{+}-\mu+\epsilon-\Delta_{22}(\omega)$,  $\Delta_{11}(\om),\ \Delta_{12}(\om),\ \Delta_{21}(\om)$ and $\Delta_{22}(\om)$ are components of the hybridisation function matrix $\hat\Delta(\om)$, and $\Sigma_{i}(\om)$ and $S_{i}(\om)$ are normal and anomalous self-energies respectively.
Now, by  doing an arithmetic averaging over Hubbard interaction, the average local Green's function is given as
\begin{equation}
\hat{G}^{CPA}(\omega)=\int d{U_{i}}\hat{G_{i}}(\omega)P_{U}({U_{i}})
\label{eq:3eq4}
\end{equation}
From equations~(\ref{eq:3eq2}, \ref{eq:3eq3}, and \ref{eq:3eq4}) $\hat{G}^{CPA}(\omega)$  is given by:,
\begin{equation}
\hat{G}^{CPA}(\omega)=(1-x)\hat{G}^{0}(\omega)+x\hat{G}^{U}(\omega)
\label{eq:3eq5}
\end{equation}
where $\hat{G}^{0}(\omega)$ and $\hat{G}^{U}(\omega)$ are given by
 \begin{equation}
\hspace*{-0cm} {\hat{G}^{0}}(\omega) =
 \left[ \begin{array}{ll}
{ \omega^{+}+\mu-\epsilon-\Delta_{11}(\omega) } & { -\Delta_{12}(\omega)} \\
 { -\Delta_{21}(\omega) }
 & {\omega^{+}-\mu+\epsilon-\Delta_{22}(\omega)}
\end{array} \right]^{-1} 
 \label{eq:3eq6}
\end{equation}
\begin{eqnarray}
\hspace*{-0cm} {\hat{G}^{U}}(\omega) =
 \left[ \begin{array}{ll}
{ \xi_{11}(\omega)-\Sigma(\omega) } & { -\Delta_{12}(\omega)-S(\omega)} \\
 { -\Delta_{21}(\omega)-S(\omega) }
 &  {\xi_{22}(\omega)+\Sigma^{*}(-\omega)}
\end{array} \right]^{-1} 
 \label{eq:3eq7}
\end{eqnarray}
To calculate $\Sigma(\omega)$ and $S(\omega)$ we employ IPTSC~\cite{3Arti} as an impurity solver. 
These self-energies are given by:
\begin{equation}
\Sigma(\omega)=-U\frac{n}{2}+A\Sigma^{(2)}(\omega)
\label{eq:3eq8}
\end{equation}
\begin{equation}
S(\omega)=-U\Phi + AS^{(2)}(\omega)
\label{eq:3eq9}
\end{equation}
\begin{equation}
\Sigma^{(2)}(\omega)=U^2\int_{-\infty}^{\infty}\prod_{i=1}^{3}d\epsilon_{i}\frac{g_{1}(\epsilon_1,\epsilon_2,\epsilon_3)N(\epsilon_1,\epsilon_2,\epsilon_3)}{\omega^{+}-\epsilon_1+\epsilon_2-\epsilon_3}
\label{eq:3eq10} 
\end{equation}
\begin{equation}
S^{(2)}(\omega)=U^2\int_{-\infty}^{\infty}\prod_{i=1}^{3}d\epsilon_{i}\frac{g_{2}(\epsilon_1,\epsilon_2,\epsilon_3)N(\epsilon_1,\epsilon_2,\epsilon_3)}{\omega^{+}-\epsilon_1+\epsilon_2-\epsilon_3} 
\label{eq:3eq11}
\end{equation}
 \begin{eqnarray}
\hspace*{0cm} {\hat{\cal{G}}^{U}}(\omega) =
 \left[ \begin{array}{ll}
{ \xi_{11}(\om)+U\frac{n}{2}} &  { -\Delta_{12}(\omega)+U\Phi} \\
 { -\Delta_{21}(\omega)+U\Phi }
 & \xi_{22}(\om)-U\frac{n}{2}
\end{array} \right]^{-1} 
 \label{eq:3eq12}
\end{eqnarray}
 \begin{equation}
\hspace*{-0cm} -\rm{Im}\frac{\hat{\cal{G}}^{U}(\omega)}{\pi} =
 \left[ \begin{array}{ll}
{ \tilde{\rho}_{11}(\omega)} & { \tilde{\rho}_{12}(\omega)} \\
 { \tilde{\rho}_{21}(\omega) }
 & {\tilde{\rho}_{22}(\omega)}
\end{array} \right]
 \label{eq:3eq13}
\end{equation}
\begin{equation}
N(\epsilon_1,\epsilon_2,\epsilon_3)=f(\epsilon_1)f(-\epsilon_2)f(\epsilon_3)+f(-\epsilon_1)f(\epsilon_2)f(-\epsilon_3)
\label{eq:3eq14}
\end{equation}
\begin{equation}
g_{1}(\epsilon_1,\epsilon_2,\epsilon_3)=\tilde{\rho}_{11}(\epsilon_1)\tilde{\rho}_{22}(\epsilon_2)\tilde{\rho}_{22}(\epsilon_3)-\tilde{\rho}_{12}(\epsilon_1)\tilde{\rho}_{22}(\epsilon_2)\tilde{\rho}_{12}(\epsilon_3)
\label{eq:3eq15}
\end{equation}
\begin{equation}
g_{2}(\epsilon_1,\epsilon_2,\epsilon_3)=\tilde{\rho}_{12}(\epsilon_1)\tilde{\rho}_{12}(\epsilon_2)\tilde{\rho}_{12}(\epsilon_3)-\tilde{\rho}_{11}(\epsilon_1)\tilde{\rho}_{12}(\epsilon_2)\tilde{\rho}_{22}(\epsilon_3)
\label{eq:3eq16}
\end{equation}
\begin{equation}
A=\frac{\frac{n}{2}(1-\frac{n}{2})-\Phi^2}{{\frac{n_{0}}{2}(1-\frac{n_{0}}{2})-\Phi_{0}^2}}
\label{eq:3eq17}
\end{equation}
where $\Phi$, $\Phi_0$, $n$ and $n_0$ are given by :,
\begin{equation}
\Phi=\int_{-\infty}^{\infty}d\omega\frac{-{\rm{Im}(G^{U}_{12}}(\omega))}{\pi}f(\omega)
\label{eq:3eq18}
\end{equation}
\begin{equation}
\Phi_{0}=\int_{-\infty}^{\infty}d\omega \tilde{\rho}_{12}(\omega) f(\omega)
\label{eq:3eq19}
\end{equation}
\begin{equation}
n=2\int_{-\infty}^{\infty}d\omega\frac{-{\rm{Im}(G^U_{11}}(\omega))}{\pi}f(\omega)
\label{eq:3eq20}
\end{equation}
\begin{equation}
n_{0}=2\int_{-\infty}^{\infty}d\omega \tilde{\rho}_{11}(\omega) f(\omega)
\label{eq:3eq21}
\end{equation}
The CPA Green's function in term of average self-energy is given by
 \begin{eqnarray}
\hspace*{0cm} \hat{G}^{CPA}(\omega) =
 \left[ \begin{array}{ll}
{ \xi_{11}(\omega)-\Sigma_{11}^{CPA}(\omega)} & { -\Delta_{12}(\omega)-\Sigma_{12}^{CPA}(\omega)} \\
 { -\Delta_{21}(\omega)-\Sigma_{21}^{CPA}(\omega)}
 & \xi_{22}(\omega)-\Sigma_{22}^{CPA}(\omega)
\end{array} \right]^{-1} 
 \label{eq:3eq22}
\end{eqnarray}
The lattice Green's function is given by:
 \begin{eqnarray}
\hspace*{0cm}  \hat{G}{(\vec{k},\omega)}=
 \left[ \begin{array}{ll}
{ \bar{\omega}_1-\epsilon{(\vec{k})}} &{ -\Sigma_{12}^{CPA}(\omega)} \\
 { -\Sigma_{21}^{CPA}(\omega)}
 & { \bar{\omega}_2 + \epsilon{(\vec{k})}}
\end{array} \right]^{-1} 
 \label{eq:3eq23}
\end{eqnarray}
where $\bar{\omega}_1=\omega^{+}-\epsilon+\mu-\Sigma_{11}^{CPA}(\omega)$ and
$\bar{\omega}_2=\omega^{+}+\epsilon-\mu-\Sigma_{22}^{CPA}(\omega)$.
Finally, the CPA self-consistency is achieved by equating the local Green's function to average impurity Green's function :,
\begin{eqnarray}
\frac{1}{N_{s}}\sum_{\vec{k}}\hat{G}{(\vec{k},\omega)}=\hat{G}^{CPA}(\omega) \nonumber \\
\hat{G}^{CPA}(\omega)=\int^{\infty}_{-\infty}d\epsilon\rho_{0}(\epsilon)\hat{G}{(\epsilon,\omega)}
\label{eq:3eq24}
\end{eqnarray}
Where, $N_{s}$ is the number of lattice sites and $\rho_{0}(\epsilon)$ is the non-interacting density of states. We have used a semi-elliptic $\rho_{0}(\epsilon)$, given by $\rho_{0}(\epsilon)=2/\pi\sqrt{(1-\epsilon^2)}$, $\epsilon\in[-1, 1]$.

\section{Numerical Algorithm}
In practice, we follow the steps outlined below to obtain the converged order parameter and spectra.
\begin{enumerate}
\item Guess a hybridization matrix $\hat\Delta(\omega)$ for interacting and non-interacting site and  $n, \Phi$ for interacting sites. In practice, we choose either a previously converged solution or the non-interacting
$\hat\Delta(\omega)$ with $n=1$ and $\Phi=1/2$.

\item Given a hybridization, occupancy and order parameter, use equation (\ref{eq:3eq12}) to calculate the host Green's function  matrix ${\hat{\cal{G}}}^U(\omega)$.

\item From equations (\ref{eq:3eq21} and \ref{eq:3eq19}), calculate pseudo-occupancy and pseudo-order parameter $n_{0}$ and $\Phi_{0}$.

\item Now, by using equations~(\ref{eq:3eq8}, \ref{eq:3eq9}, \ref{eq:3eq10}, \ref{eq:3eq11}, and \ref{eq:3eq17}), calculate the regular and anomalous self-energies $\Sigma(\omega)$ and $S(\omega)$.

\item Then,  by using equations~(\ref{eq:3eq6}, \ref{eq:3eq7}, \ref{eq:3eq8} and \ref{eq:3eq9} ), calculate impurity Green's function $\hat G^0(\omega)$, $\hat G^U(\omega)$, $n$ and $\Phi$ .

\item The disorder-averaged Green's function, $\hat{G}^{\CPA}(\omega)$ is obtained using equation~(\ref{eq:3eq5}).

\item We consider the AHM on Bethe lattice of infinite connectivity at half filling, which is 
achieved by setting $\mu=0, \epsilon=0$. For Bethe lattice the self-consistency condition is simply given by :,
\begin{equation}
\hat{\Delta}(\omega)=\frac{t^2\sigma_{z}\hat{G}^{\CPA}(\omega)\sigma_{z}}{4}
\label{eq:3eq25}
\end{equation}
where $\sigma_{z}$ is z component of Pauli's matrix. Using equation~(\ref{eq:3eq25}), a new hybridization matrix $\hat\Delta(\om)$ is obtained. 

\item If the hybridization matrix $\hat\Delta(\omega)$ from step 7 and $n,\Phi$
 from step 5 are equal (within a desired numerical tolerance) to the guess  hybridization matrix $\hat\Delta(\omega)$, $n$ and $\Phi$ from step 1, then the iterations may be stopped, else the iterations continue until the desired convergence is achieved. 

\end{enumerate}
The results obtained using the above-mentioned procedure will be denoted as
IPTSC. We have also carried out mean-field calculations by ‘turning off’ the dynamical
self-energies in equations~\ref{eq:3eq8} and \ref{eq:3eq9}. These results will be denoted as BdGMF.
\section{Results and Discussion}
In this paper, we have considered $1-x$ fraction of the sites to be
non-interacting ($U=0$), and $x$ fraction to be interacting ($U\neq0$). The
unit of energy is the hopping integral $t=1$. We have done all the 
calculations at half filling ($\langle n\rangle =1$),  which is fixed by 
taking $\mu=\epsilon=0$. 
\subsection{Spectral functions}
\begin{figure}[]
\centering
\includegraphics[scale=0.3,clip]{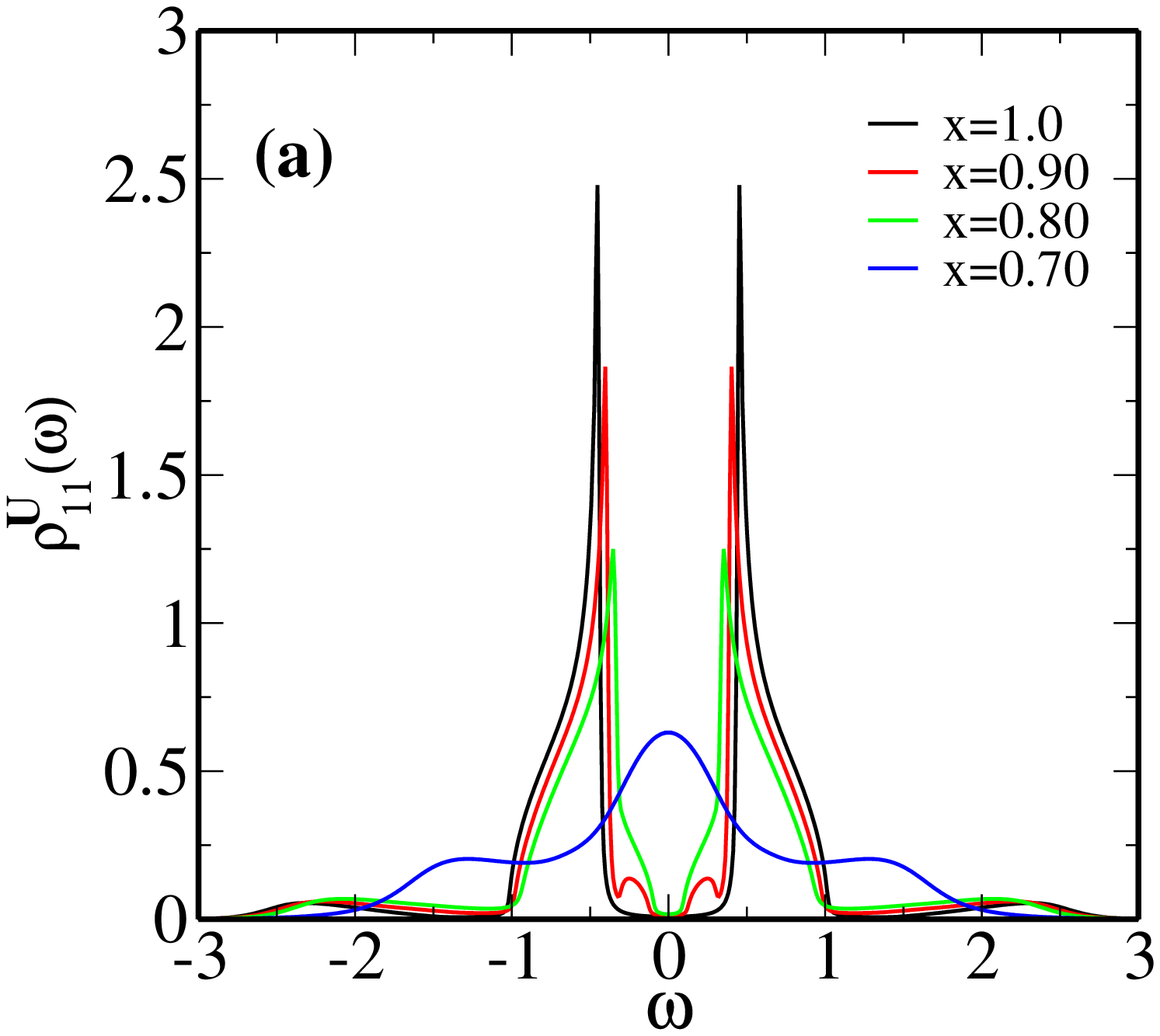}
\includegraphics[scale=0.3,clip]{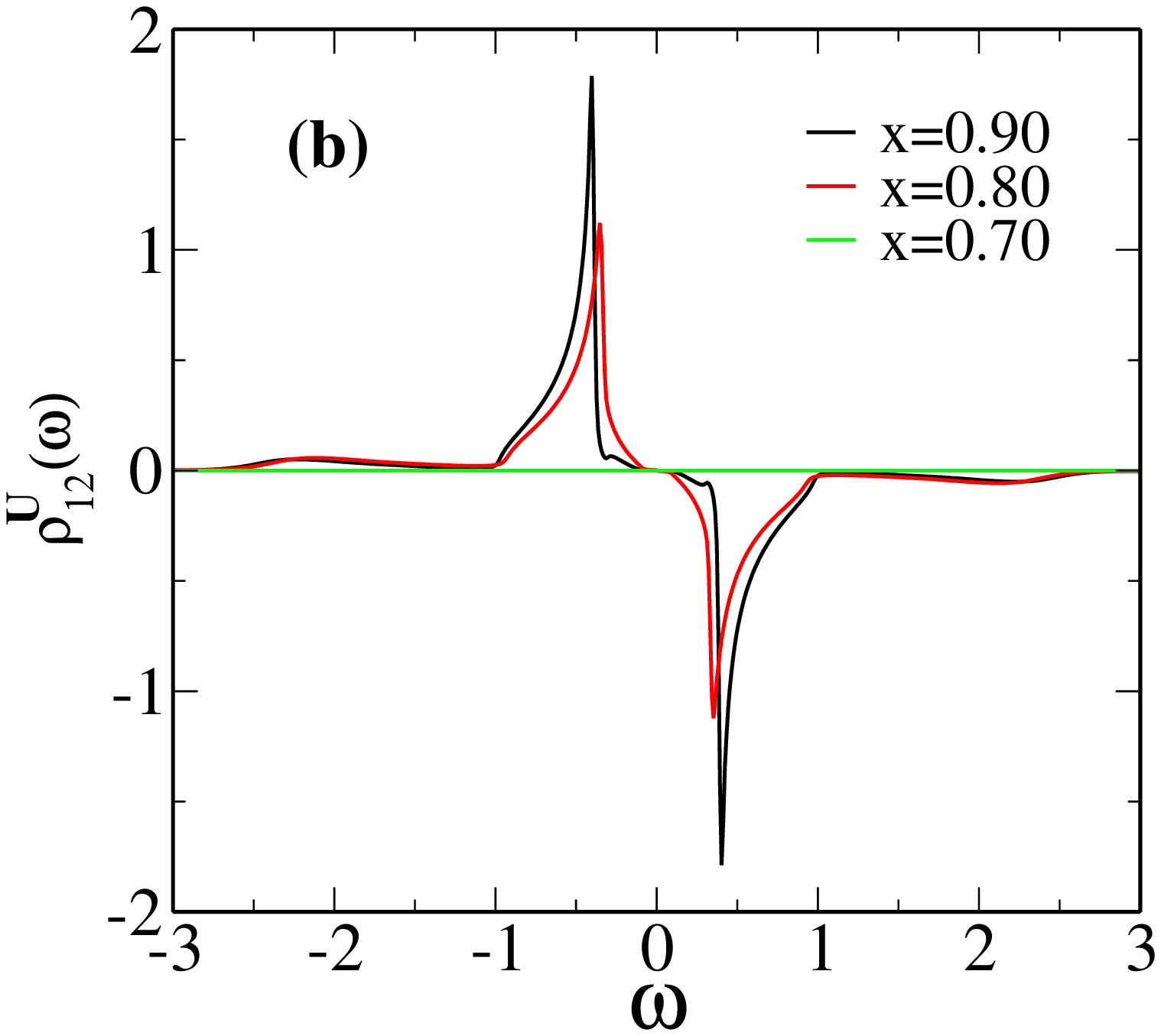}
\includegraphics[scale=0.3,clip]{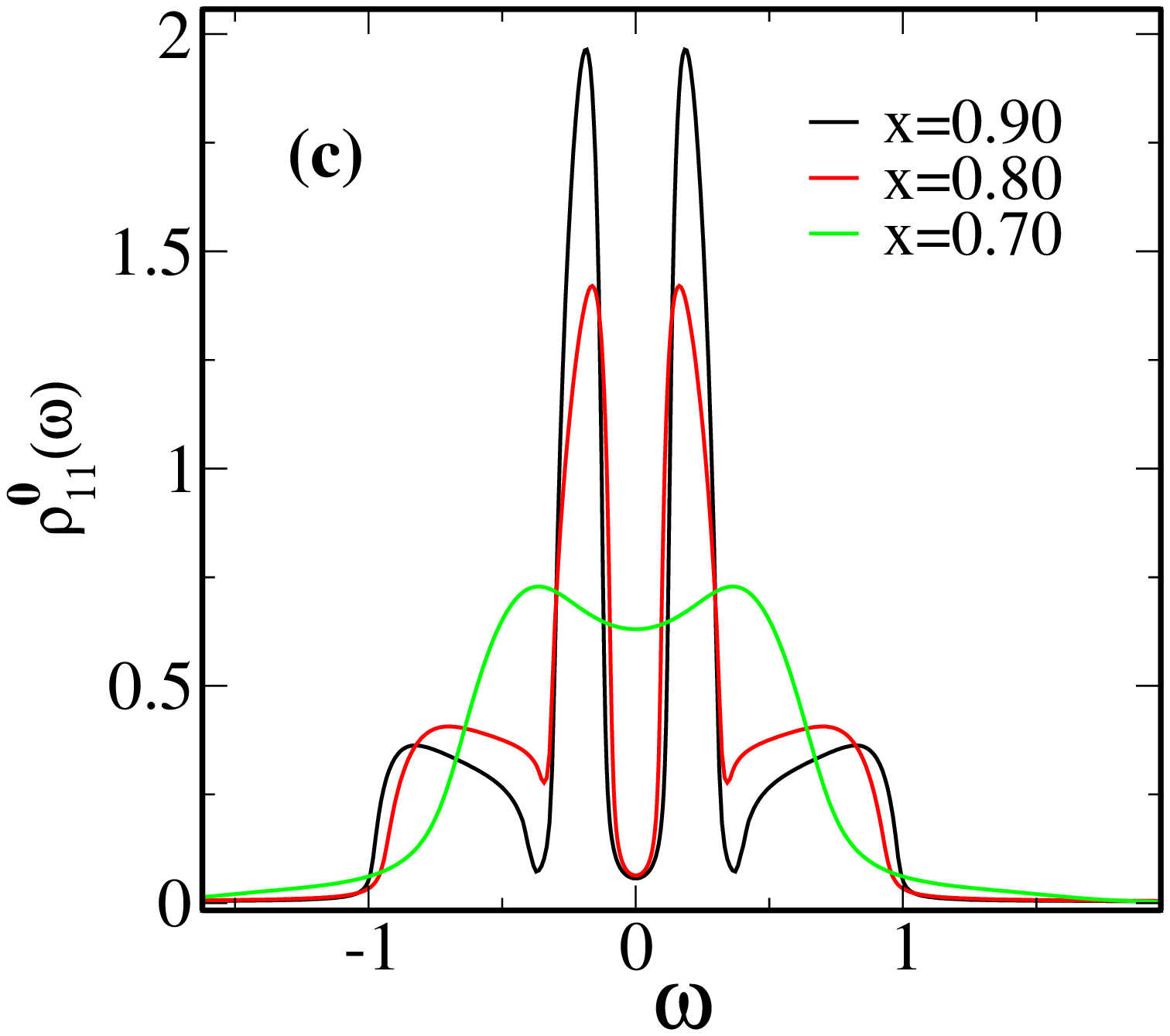}
\includegraphics[scale=0.3,clip]{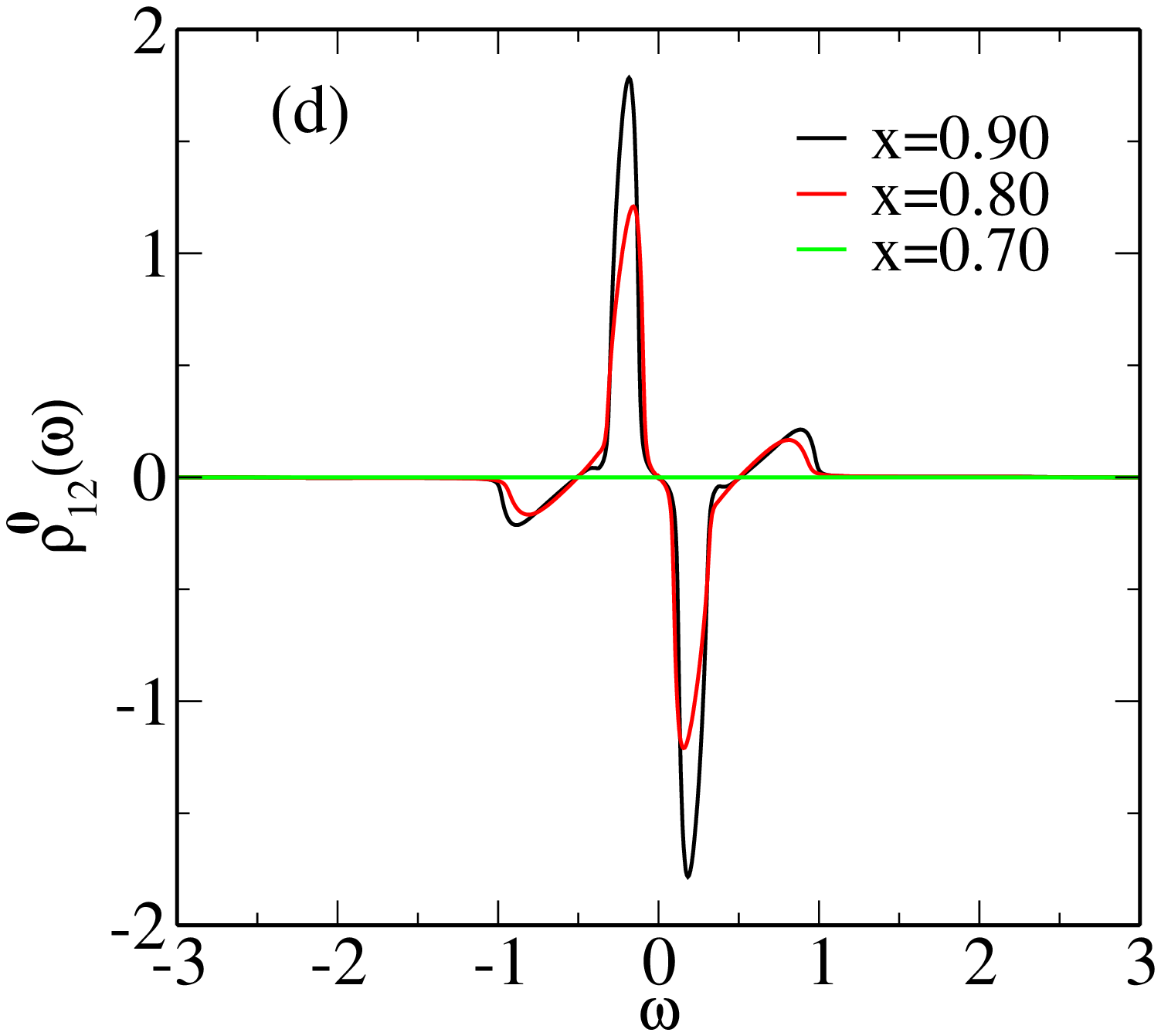}
\caption{Diagonal (a,c) and off-diagonal (b,d) spectral function as a function of frequency for interacting and non-interacting site respectively at U=1.5.}
\label{fig:3fig1}
\end{figure}

In figure~\ref{fig:3fig1}, the  11 and 12 components of local spectral functions as a function of frequency are shown for different values of $x$ at $U=1.5$. The panel (a) and (b) represent the 11 and 12 components of spectral function of interacting site respectively.   $\rho^U_{11}(\omega)$, $\rho^U_{12}(\omega)$, $\rho^0_{11}(\omega)$, and $\rho^0_{12}(\omega)$  represent the 11 and 12 components of local spectral function of the interacting   and non-interacting sites respectively.  
For a $U=1.5$, we find that, beyond a critical value of $x\sim0.70$ the  $\rho^U_{11}(\omega)$ becomes gapped, and the gap increases with increasing $x$. The spectral function has coherence peaks at the gap edges, and the weight
 of the coherence peak increases with increasing $x$. Concomitantly, beyond $x\sim0.70$, the
off-diagonal spectrum,  $\rho^U_{12}(\omega)$ develops finite spectral weight, which increases with increasing $x$. This also implies that the local superconducting order parameter,
$\Phi$, given by integration of $\rho^U_{12}(\omega)$ upto the Fermi level, increases with increasing $x$. Thus we conclude that
the spectral gap of figure~\ref{fig:3fig1}(a) is superconducting in nature. 
We find that the non-interacting sites also develop superconductivity.
This is seen from the evolution of the local spectral functions corresponding to the non-interacting sites, which are shown in the two lower panel of figure~\ref{fig:3fig1}. This is natural because the host in which
the interacting and non-interacting sites are embedded, characterized by
$\hat\Delta(\om)$, becomes superconducting.

\begin{figure}[]
\centering
\includegraphics[scale=0.35]{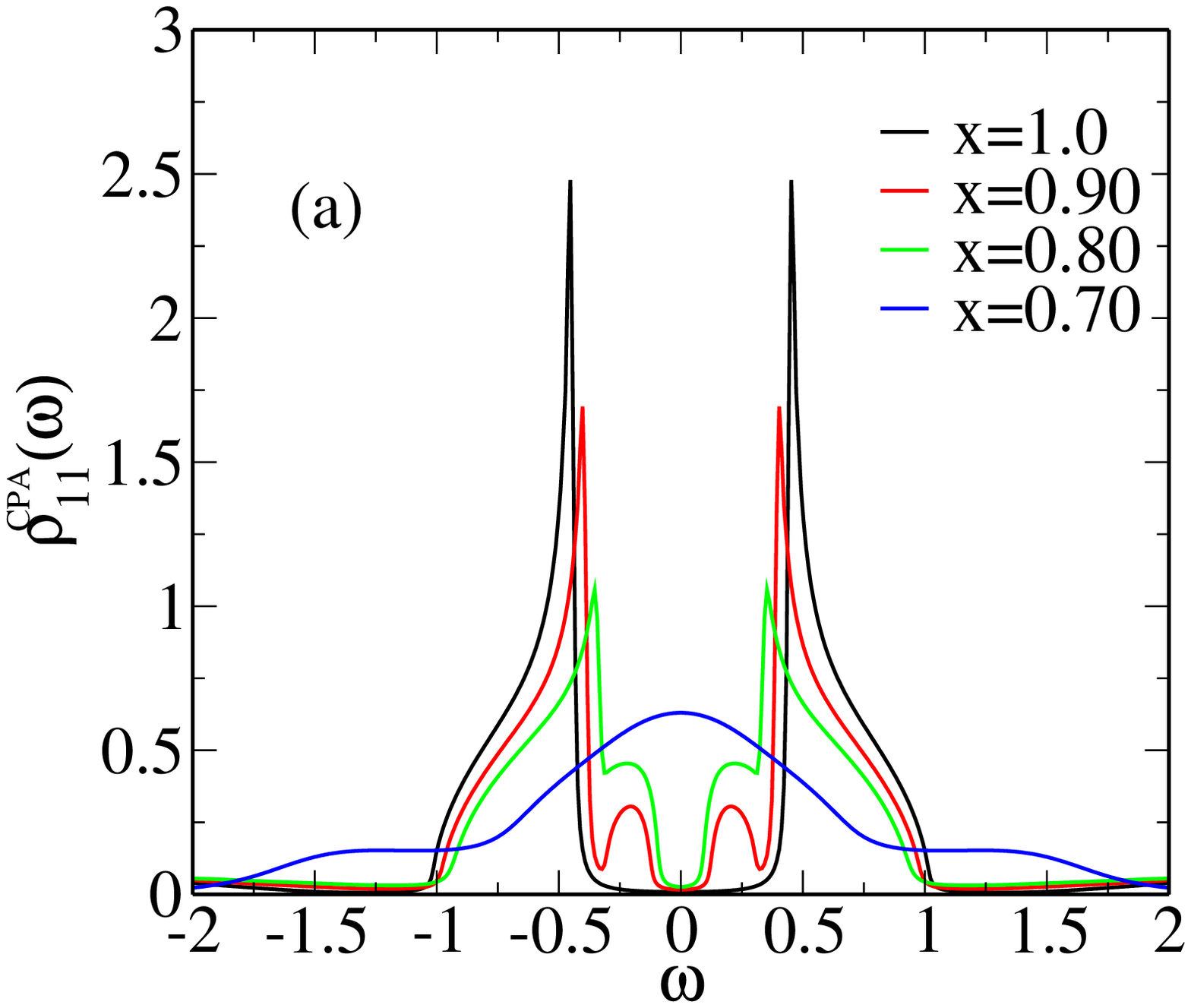}
\includegraphics[scale=0.35]{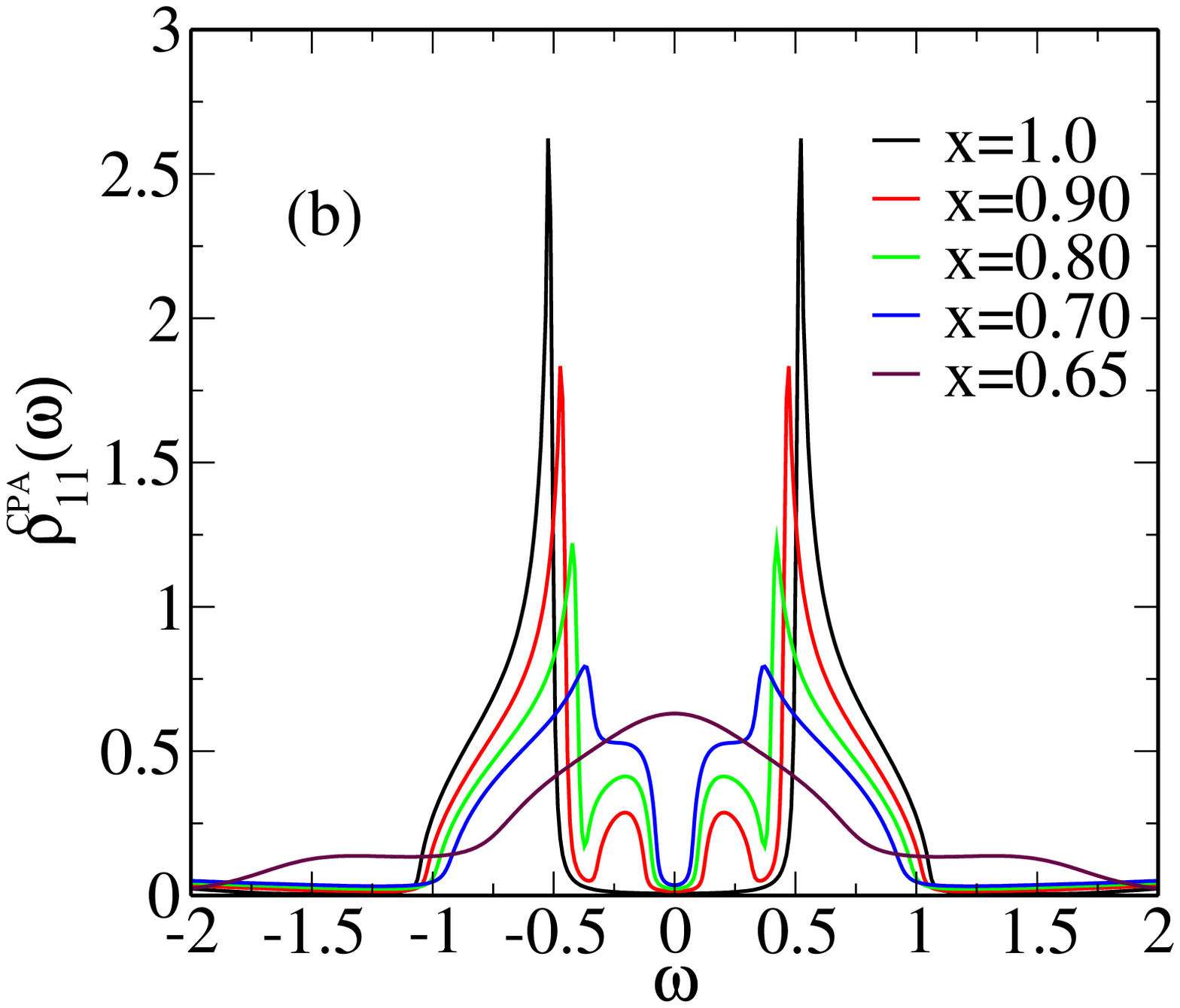}
\includegraphics[scale=0.35]{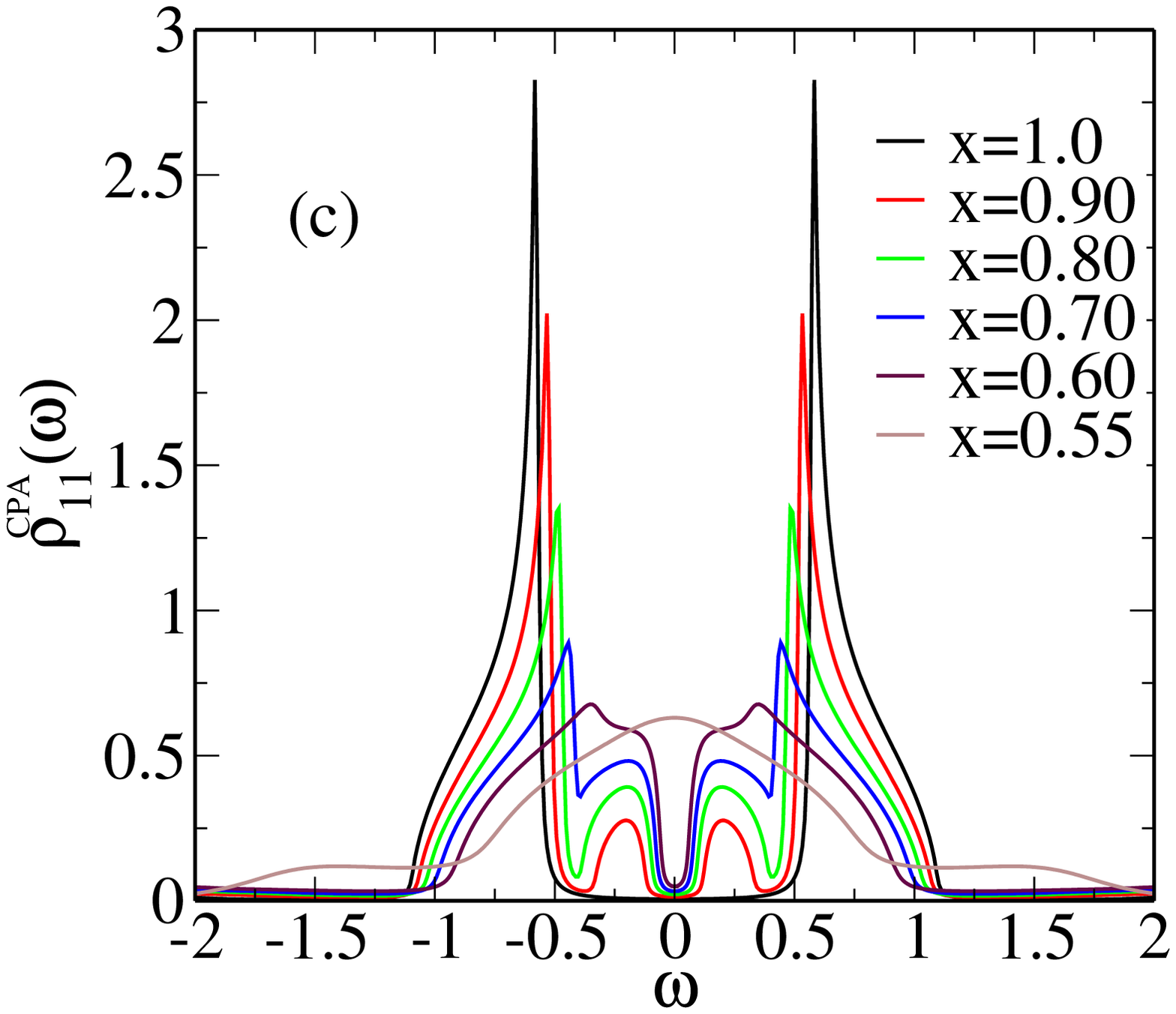}
\includegraphics[scale=0.35]{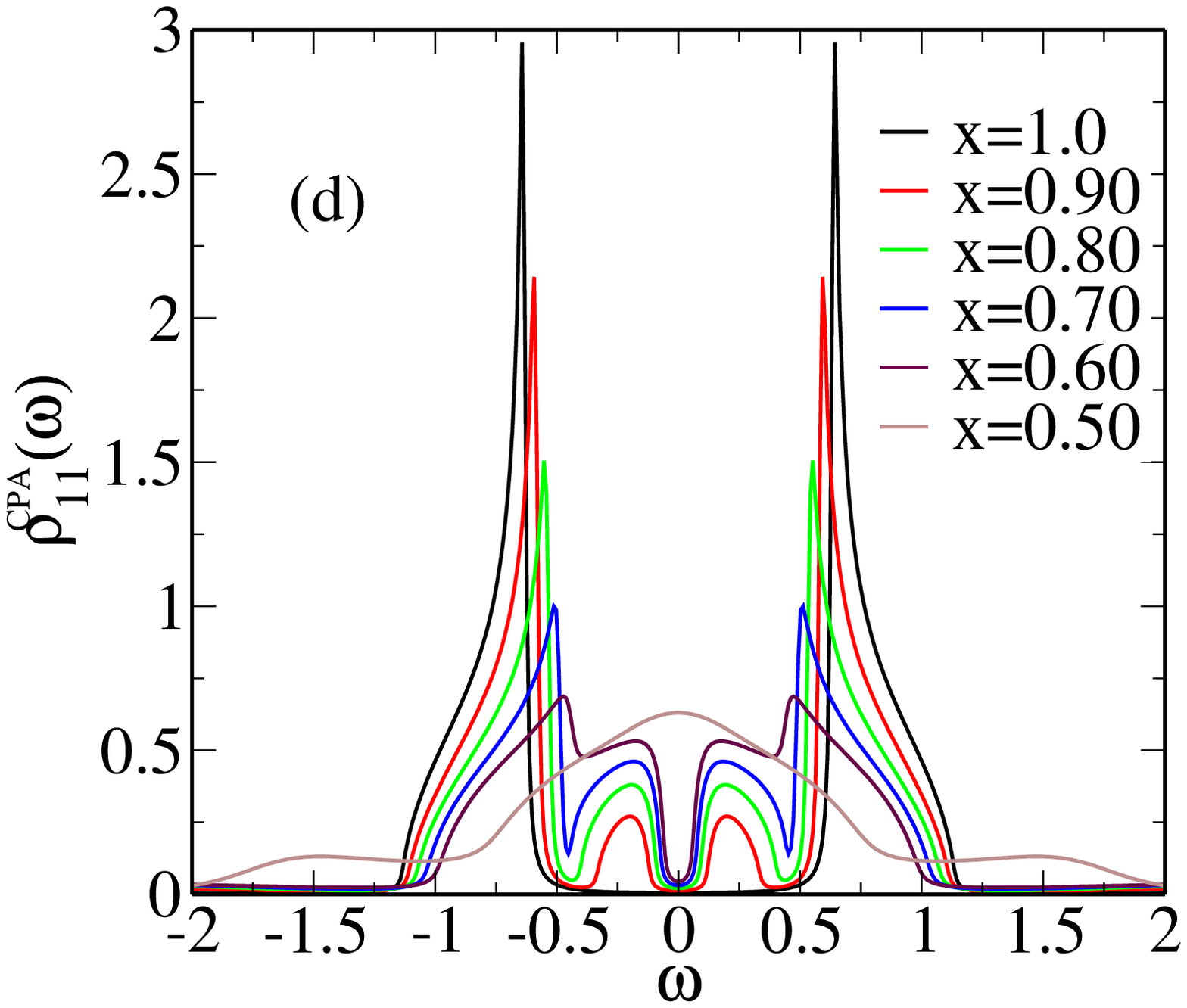}
\caption{ Diagonal component of CPA spectral function  as a function of frequency for (a) U=1.5 (b) U=1.6 (c) U=1.7 (d) U=1.8.}
\label{fig:3fig2}
\end{figure}
In  figure~\ref{fig:3fig2} the 11 component of disorder averaged spectral function is shown for various values of $x$ computed with $U=1.5, 1.6, 1.7$ and, $1.8$.
It is seen all the spectra are gapped at $x=1.0$. With increasing $x$, the gap
increases, and finally a   metal to superconductor transition occurs at a critical 
$x_c$. The nature of this transition and the dependence of $x_c$ on $U$ will
be discussed next.

\subsection{Metal-superconductor transition}
\begin{figure}[]
\centering
\includegraphics[scale=0.45,clip]{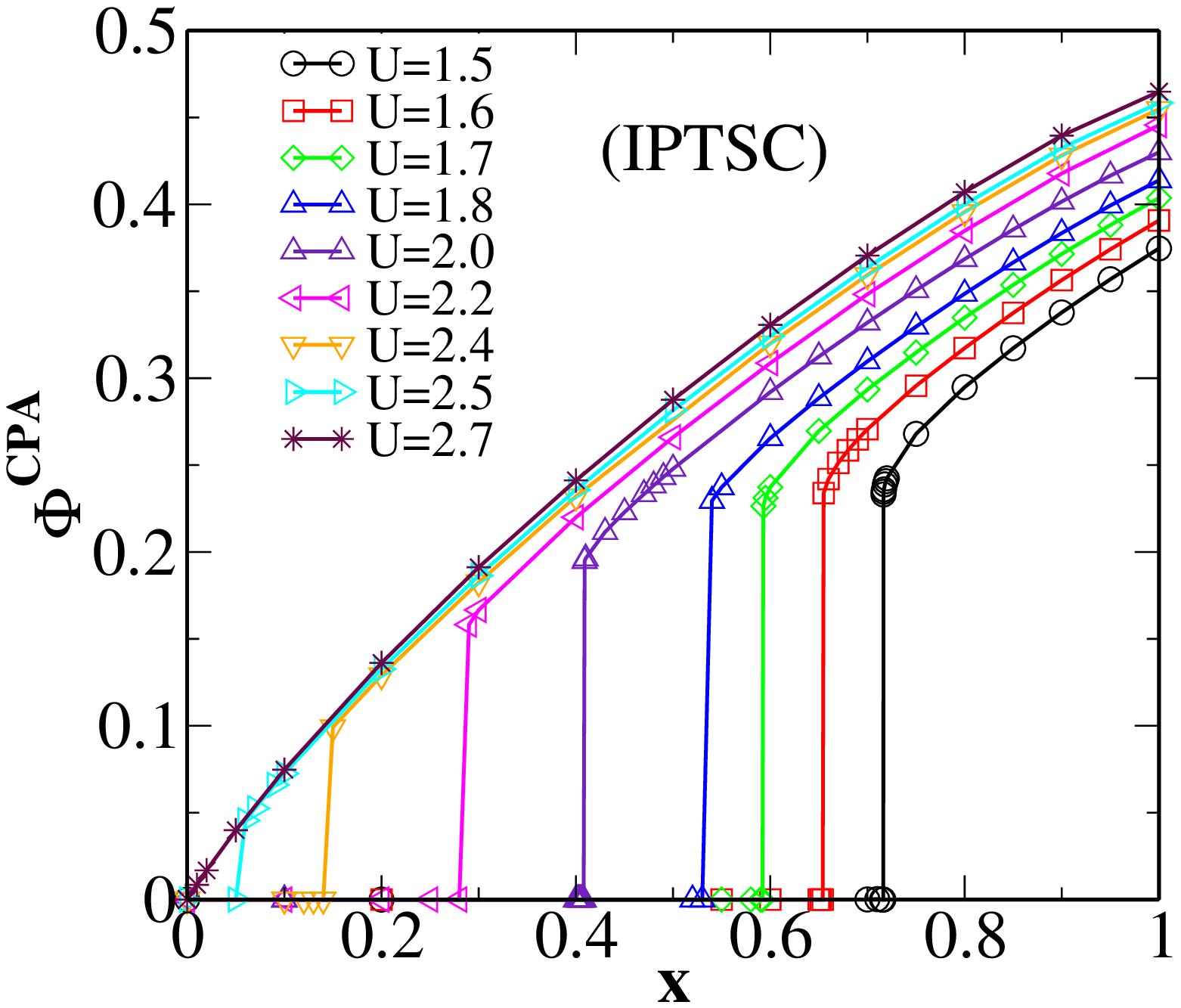}
\includegraphics[scale=0.45,clip]{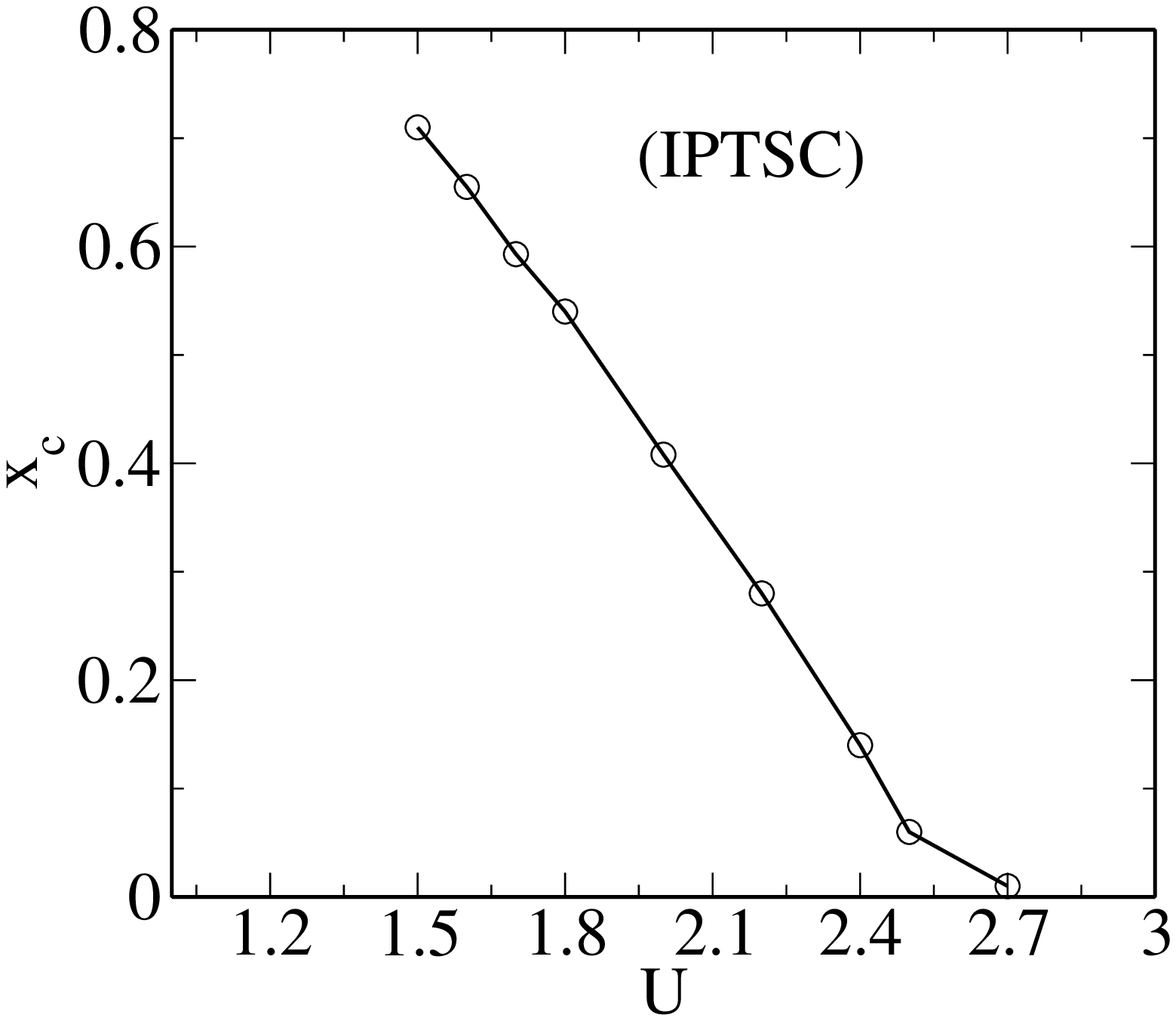}
\caption{Upper panel: Disorder averaged superconducting order $( \Phi^{CPA} )$ as a function of $x$ for different values of $U$. Lower Panel : The critical
fraction of interacting sites, $x_c$  vs $U$.}
\label{fig:3fig3}
\end{figure}

The results in the previous sections suggest the following scenario: For $x=0$,
there are no sites with $-U$, and the system is a metal. With increasing $x$,
the system develops superconductivity beyond a critical $x_c$.  The disorder
 averaged superconducting order parameter $(\Phi^{CPA})$, calculated by using the expression, 
\begin{equation}
\Phi^{CPA}=\int_{-\infty}^{\infty}d\omega\frac{-\rm{Im}(G^{CPA}_{12}(\omega))f(\omega)}{\pi}\,,
\label{eq:3eq26}
\end{equation}
is shown in upper panel of figure~\ref{fig:3fig3}  as a function of $x$ for different values of $U$. 
For all $U\lesssim 2.7$, a finite $x_c$ is needed
before the superconducting order
develops. The transition from metal to superconductor is seen to be
first order. The critical $x_c$ decreases sharply with increasing $U$ as seen in the lower panel.   For all $U\geqslant 2.7$, the transition becomes continuous
and the critical $x_c$ needed to generate superconductivity is nearly zero.
This indicates that the interaction stabilises the superconducting phase despite the presence of disorder. The $x_c\rightarrow 0$ for $U\gtrsim 2.7$ is most likely
an artefact of the infinite coordination number within dynamical mean field theory. Including non-local dynamical correlations through quantum cluster  
theories~\cite{qct} would most likely lead to an asymptotic decay of $x_c$ with increasing $U$.
\subsection{Comparison of IPTSC and BdGMF results}
\begin{figure}[ht]
\centering
\includegraphics[scale=0.38,clip]{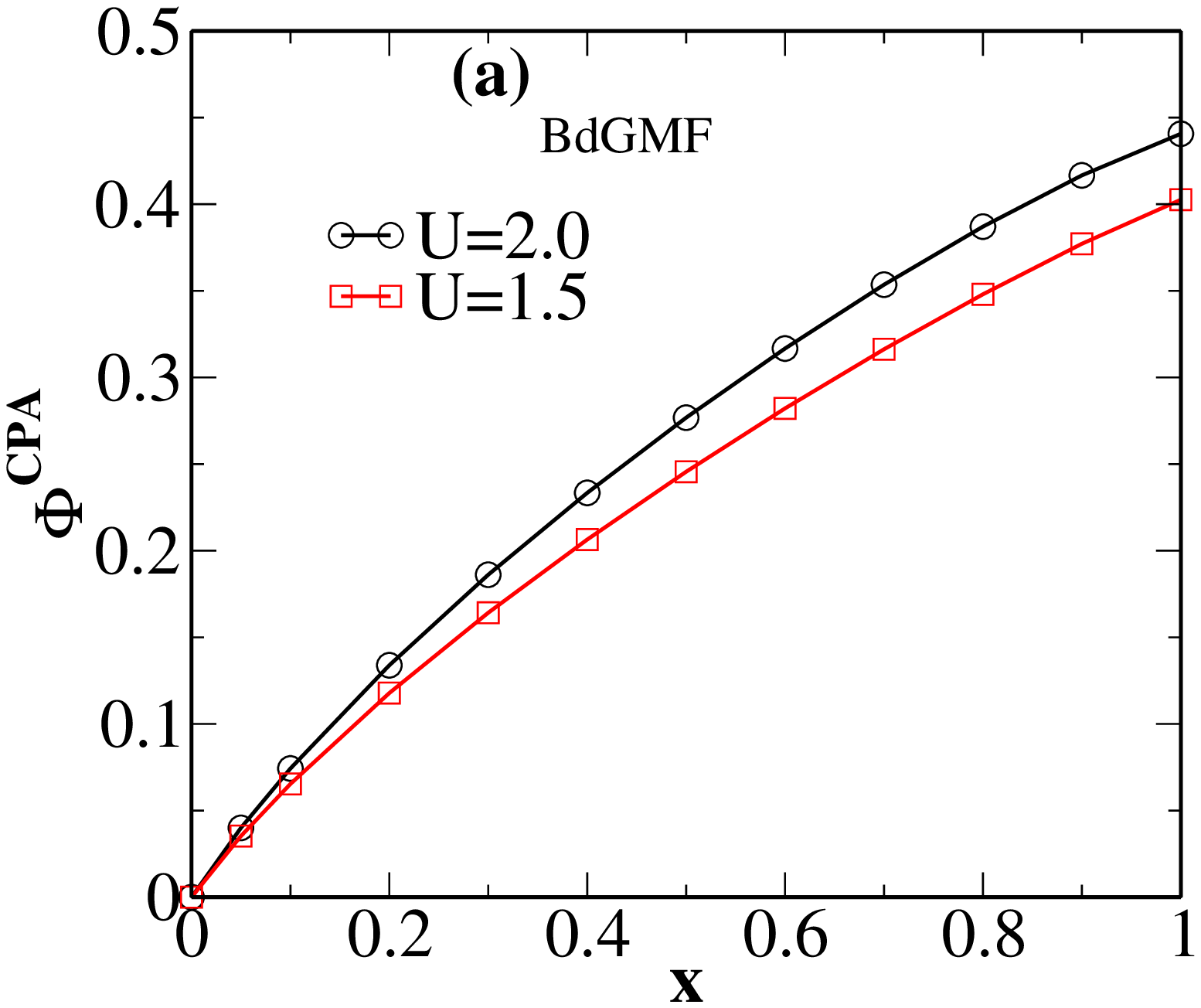}
\includegraphics[scale=0.38,clip]{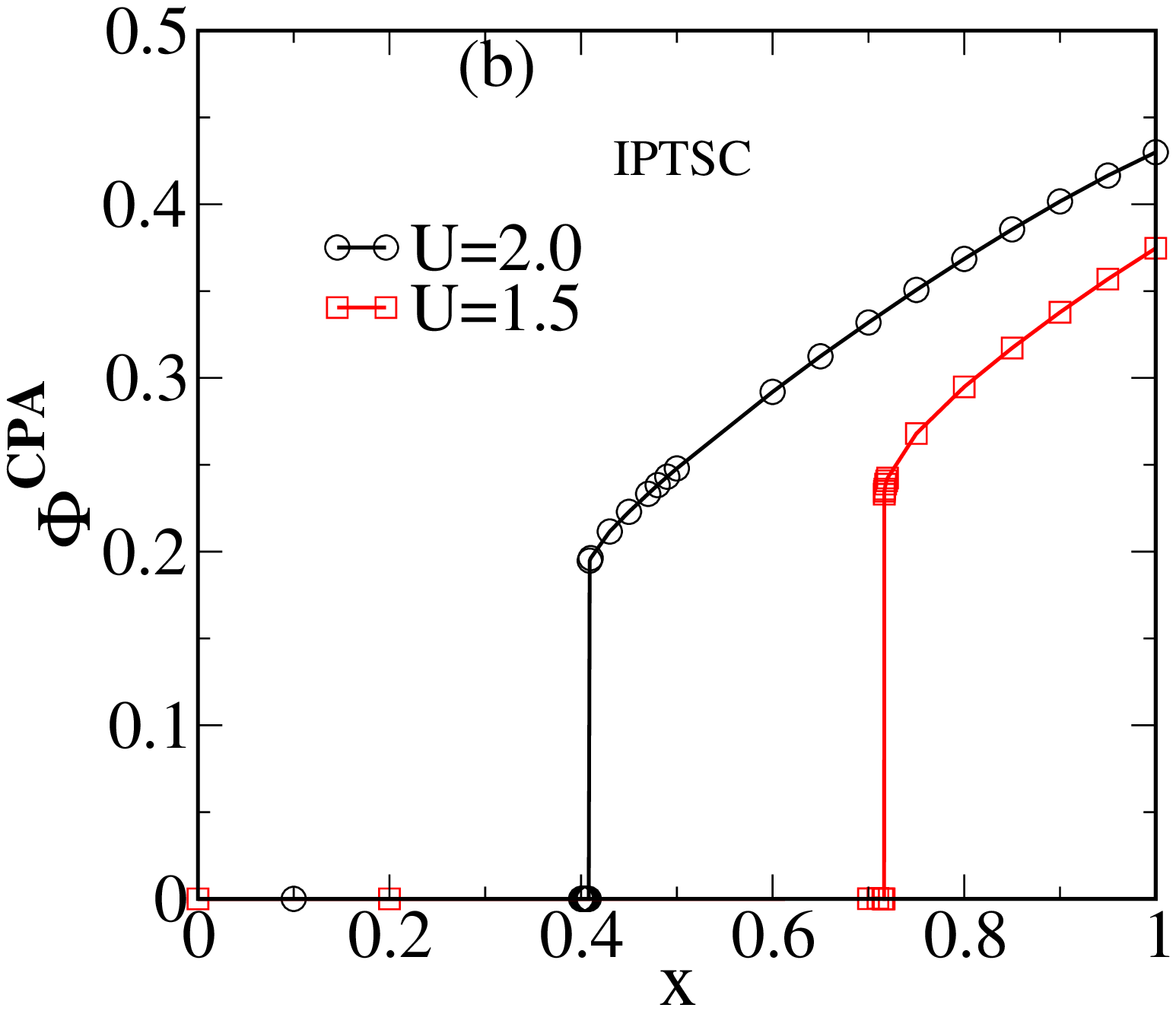}
\caption{ The disorder-averaged superconducting order parameter, 
$\Phi^{CPA}$, as a function of disorder, $x$ computed
within BdGMF (upper panel) and IPTSC (lower panel).}
\label{fig:3fig5}
\end{figure}

We would like to understand the precise effect of incorporating dynamics
beyond static mean field solutions. Hence we compare a few representative
 results from IPTSC with those from BdGMF.
  Figure~\ref{fig:3fig5} depicts this comparison.   
The $\Phi^{CPA}$ computed with BdGMF and IPTSC are shown in panels (a) and 
(b) respectively. As seen, the BdGMF $\Phi^{CPA}$  increases continuously 
with increasing $x$,  while the DMFT calculation, using IPTSC
as the solver, shows a first order transition. Thus, incorporating dynamics
changes the qualitative nature of the metal-superconductor transition.
\section{Conclusions }
In this paper, a disordered attractive Hubbard model with 
spatially random interaction sites is investigated by combining DMFT, CPA
and IPTSC as an impurity solver at half filling. We have computed local quantities such as diagonal and off diagonal spectral function for different values of U and $x$. By using local  off diagonal spectral functions, we have computed superconducting order parameter. We find a doping (disorder) induced metal to superconductor transition.
The transition is first order for low interaction strengths, but becomes continuous for $U> U_c$. The critical disorder needed to
achieve superconductivity becomes nearly zero beyond $U>U_c$.
The experiments on Tl doped PbTe, mentioned in the introduction, 
show that~\cite{3Matsushita1,3Matsushita2}
superconductivity is induced with just 0.3\% concentration of Tl. From our
 figure~\ref{fig:3fig3}, we would then conclude that the local attraction strength at the Tl-sites is quite larger than the bandwidth. 
To highlight the effects of dynamical fluctuations beyond 
static mean-field, we have calculated the superconducting order parameter in IPTSC and BdGMF frameworks. In parallel to the IPTSC scenario described above, the BdGMF approach shows a metal to superconductor  transition, however
the transition is always continuous for all attractive interaction strengths.

\begin{acknowledgment}
Authors thank CSIR, India and JNCASR, India for funding the research.
\end{acknowledgment}

\end{document}